\documentstyle[prb,aps,epsf]{revtex}

\begin{document}
\draft

\preprint{UdeM.\ Rep.\ No.\ PMC/LJL/97-04}

\twocolumn[\hsize\textwidth\columnwidth\hsize\csname @twocolumnfalse\endcsname

\title
{\bf Island morphology and adatom self-diffusion on Pt(111)}

\author{Ghyslain Boisvert\cite{byline1} and Laurent J. Lewis\cite{byline2}}
\address{
D{\'e}partement de Physique et Groupe de Recherche en Physique et Technologie des
Couches Minces (GCM), Universit{\'e} de Montr{\'e}al, Case Postale 6128, Succursale
Centre-Ville, Montr{\'e}al, Qu{\'e}bec, Canada H3C 3J7
}

\author{Matthias Scheffler}
\address{
Fritz-Haber-Institut der Max-Planck-Gesellschaft, Faradayweg 4-6, D-14195
Berlin-Dahlem, Germany
}

\maketitle

\begin{center}
{\bf Submitted to Physical Review B} \\ \today
\end{center}

\begin{abstract}

The results of a density-functional-theory study of the formation energies of
(100)- and (111)-faceted steps on the Pt(111) surface, as well as of the
barrier for diffusion of an adatom on the flat surface, are presented. The
step formation energies are found to be in a ratio of 0.88 in favour of the
(111)-faceted step, in excellent agreement with experiment; the equilibrium
shape of islands should therefore clearly be non-hexagonal. The origin of the
difference between the two steps is discussed in terms of the release of
stress at the surface through relaxation. For the diffusion barrier, we also
find relaxation to be important, leading to a 20$\%$ decrease of its energy.
The value we obtain, 0.33 eV, however remains higher than available
experimental data; possible reasons for this discrepancy are discussed. We
find the ratio of step formation energies and the diffusion barrier to be the
same whether using the local-density approximation or the
generalized-gradient approximation for the exchange-and-correlation energy.

\end{abstract}


\vskip2pc
]

\narrowtext

\section{Introduction}\label{intro}

Detailed knowledge of surface properties is important to understanding a wide
variety of phenomena such as catalysis, surface reactivity, growth, etc. Of
particular importance are such quantities as the step formation energy (SFE),
which determines the equilibrium shape of islands on flat terraces, and the
potential energy surface seen by, e.g., an adatom, which provides information
on the preferred sites for adsorption and the kinetics of diffusion. We
present here a detailed study of these properties for the (111) surface of
platinum within the framework of density-functional theory (DFT).\cite{kohn}

\begin{figure}
\vspace*{-0.25in}
\epsfxsize=3.in \epsfbox{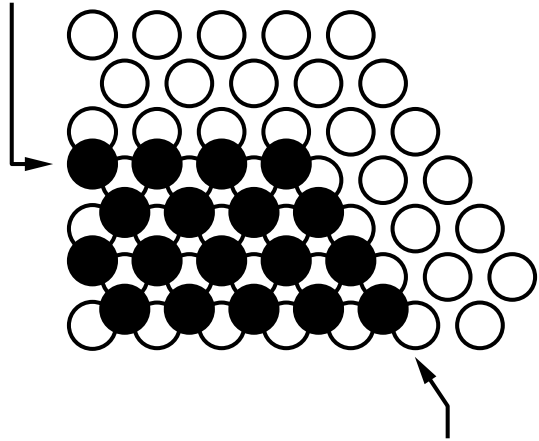}
%\vspace{-0.25in}
\caption{
Top view of an island (filled circles) on a fcc(111) surface indicating the
two possible step geometries.
\label{steps}
}
\end{figure}

The (111) surface of fcc metals, and in particular Pt, is of interest for (at
least) two reasons. First, as displayed in Fig.\ \ref{steps}, it exhibits two
possible step geometries, named according to the micro-facet that step-edge
atoms form with atoms in the layer underneath, namely a (100)-faceted step,
where an edge atom has a single near neighbor at the base of the step (i.e.,
the micro-facet constitutes a square lattice) and a (111)-faceted step, where
each atom has two neighbors (triangular lattice). The only difference between
the two geometries, as far as nearest-neighbors are concerned, is that atoms
at the base of the step have a coordination of 10 for the (100)-faceted step
and 11 for the (111)-faceted step. The two steps are evidently very similar
and the formation energies are thus {\em expected} to be comparable, i.e.,
the equilibrium island shape should be very nearly hexagonal, with the sides
consisting of, alternately, (100)- and (111)-faceted steps. This has indeed
been observed in the case of Ir\cite{fu} and Ag.\cite{morgenstern} For Pt,
however, scanning-tunneling microscopy (STM) reveals, rather, a strong
preference for (111)-faceted steps --- 0.87$\pm$0.02 as measured by the ratio
of step formation energies per unit length.\cite{michely} On the theory side,
DFT calculations in the local-density approximation\cite{ceperley} (LDA) by
Stumpf and Scheffler predict this behaviour in the case of Al, with a ratio
of SFE of 0.93;\cite{stumpf1} this has not yet been confirmed experimentally.
For Pt, in contrast, corresponding calculations by Feibelman failed to
reproduce the experimental results, leading to essentially equal SFE for the
two types of steps.\cite{feibelman0}

\begin{figure}
%\vspace*{-0.25in}
\epsfxsize=3.in \epsfbox{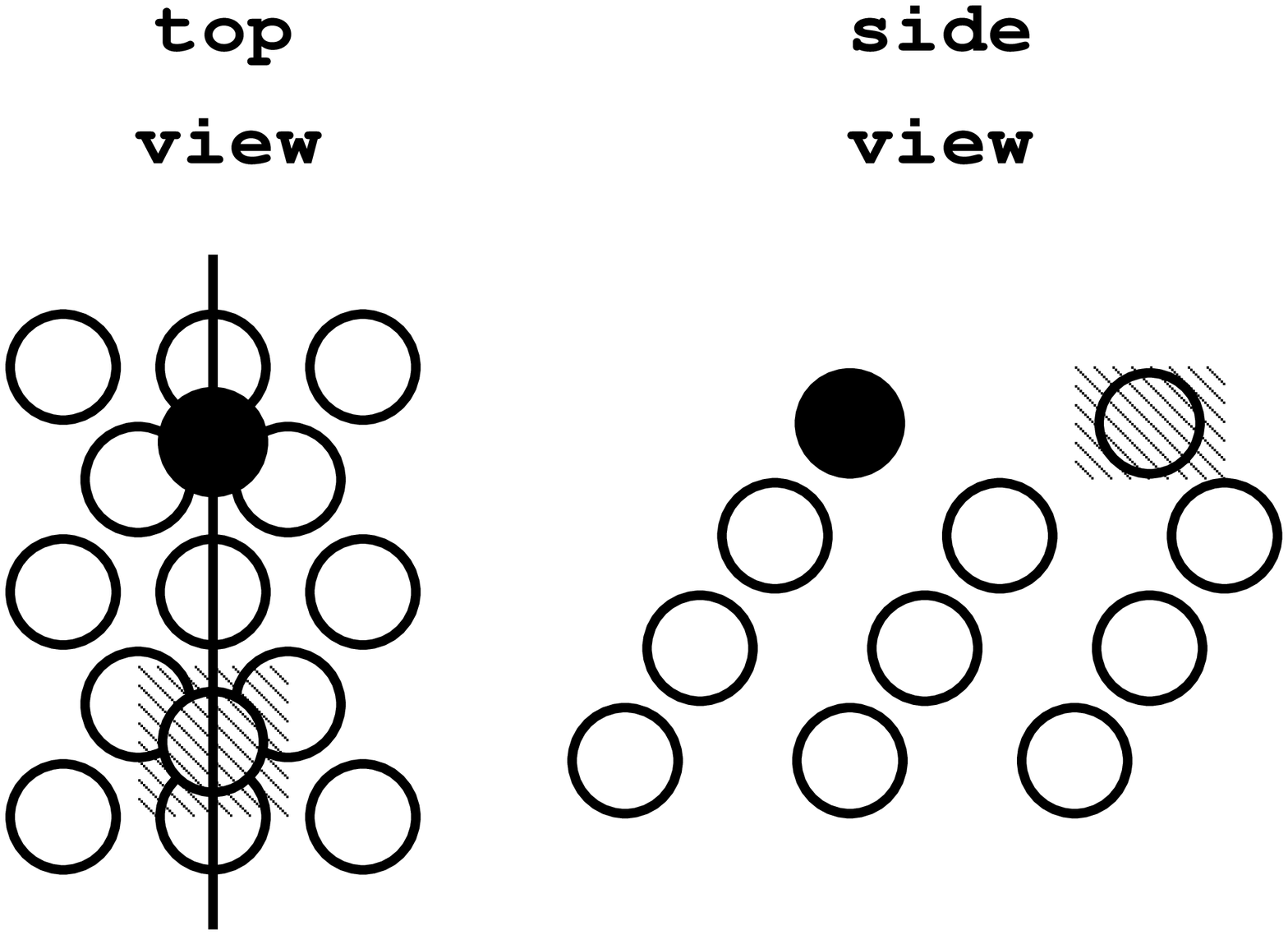}
\vspace{-0.5in}
\caption{
Top and side view of the two different adsorption sites on a fcc (111)
surface: fcc site (filled circle) and hcp site (dashed circle). The side view
corresponds to the plane indicated by the straight line in the top view.
\label{sites}
}
\end{figure}

A second feature that makes the fcc(111) surface interesting is that it
possesses two different adsorption sites: the fcc (normal) site, where an
adatom sits in a position appropriate to the stacking of atomic planes in a
perfect fcc crystal, and the hcp (stacking fault) site, corresponding to an
hcp stacking. Both sites have three nearest neighbors, the difference between
the two lying in the second-layer arrangement, as illustrated in Fig.\
\ref{sites}. It has been shown that, for transition metals, the preferred
site for adsorption depends on the filling of the $d$
band.\cite{mortensen,piveteau} For Pt, DFT-LDA calculations predict the fcc
site to be much more favourable,\cite{mortensen,feibelman1} by 0.12--0.18 eV,
depending on the state of relaxation of the substrate. This is the largest
difference observed (so far) for late transition metals and noble
metals.\cite{mortensen,boisvert1,wang} Experimentally, also, there is
evidence that the fcc site is preferred over the hcp site in
Pt;\cite{feibelman1,golzhauser} while a precise numerical value cannot be
inferred from the measurements, it is estimated that the difference should be
{\em at least} 0.06 eV.\cite{golzhauser} However, despite this apparent
agreement, the DFT calculations for Pt(111)\cite{mortensen,feibelman1} have
been unable to reproduce correctly the diffusion barrier --- 0.38--0.41
versus $\sim$0.26 eV from experiment.\cite{feibelman1,bott} Since it is of
primary importance to have reliable and accurate energy barriers in order to
predict growth (see for example Ref. \onlinecite{ruggerone}), it is essential
that this discrepancy be resolved.

Clearly, a quantitative picture of the surface properties of Pt(111) is still
missing. In order to address this problem, we present here the results of
extensive {\em ab-initio} total-energy calculations of the formation energies
of the two kinds of steps, as well as of the energetics of adatom adsorption
and diffusion. The calculations have been carried out within the LDA, but we
have also carried out a series of calculations in the generalized-gradient
approximation (GGA)\cite{perdew} so as to assess the applicability of the LDA
to this system. We find the LDA to offer a better description (compared to
experiment) of the lattice constant of bulk Pt than the GGA, in agreement
with previous calculations,\cite{ozolins,khein} as well as some properties of
the clean (111) surface (surface relaxation and work function). Within
numerical accuracy, however, we observe no sizeable effect on the ratio of
SFE and on the energetics of adatom adsorption and diffusion.

As discussed in more detail below, the electronic wave-functions were
expanded in plane waves, which enables us to deal easily, and completely,
with the effect of relaxation on adsorption and adatom diffusion. In the
calculations of Refs.\ \onlinecite{mortensen} and \onlinecite{feibelman1},
only nearest-neighbor relaxation was included, at best. We observe that
proper account of atomic relaxation leads to a significant decrease, by
$\sim$20$\%$, of the value of the diffusion barrier on the flat (111)
surface. Our estimate for this quantity, while closer to
experiment\cite{feibelman1,bott} than previous
calculations,\cite{mortensen,feibelman1} however remains high --- 0.33 versus
$\sim0.26$ eV; possible reasons for this discrepancy will be discussed. In
contrast, the difference in energy between the two adsorption sites increases
upon relaxing, from 0.10 to 0.17 eV.

One other important advantage of plane waves is that, contrary to the
Gaussian orbitals used in Ref.\ \onlinecite{feibelman0}, they are independent
of the atomic positions and should therefore provide a more adequate
description of the subtle difference in SFE expected here. In agreement with
experiment, our highly-converged calculations indicate a clear preference for
(111)-faceted steps over (100)-faceted steps, the SFE being in a ratio of
0.88 (versus about 0.87 experimentally).\cite{michely} Our calculations,
further, provide a simple explanation for the origin of the energy difference
between the two steps in terms of the release of surface stress through
relaxation. Before discussing our results in detail, we give a brief
description of our computational approach.

\section{Computational details}\label{model}

As already noted above, the calculations reported here were performed within
the framework of density-functional theory,\cite{kohn} using both the
LDA\cite{ceperley} and the GGA\cite{perdew} for the exchange-and-correlation
energy. The ion cores were approximated by pseudopotentials with $5d$
electrons treated as valence states. The pseudopotentials were generated
using the semi-relativistic scheme of Troullier and Martins\cite{troullier}
and expressed in the Kleinman-Bylander form using the $s$ component as the
local one.\cite{kleinman,gonze,fuchs} The electronic wave-functions were
represented using a plane-wave basis set with kinetic energy up to 40 Ry in
the LDA and 45 Ry in the GGA. To improve convergence, the electronic states
were occupied according to a Fermi distribution with $k_BT_{\rm el} = 0.1$ eV
and the total energies obtained by extrapolating to zero electronic
temperature. For similar reasons, the calculations were initiated using
wave-functions obtained from the self-consistent solution of the Kohn-Sham
Hamiltonian in a mixed basis set composed of pseudo-atomic orbitals and plane
waves cut off at 4 Ry.\cite{kley} The minimization of the energy with respect
to the electronic degrees of freedom was done using an iterative
procedure.\cite{stumpf} After achieving electronic convergence, the atoms
were moved according to a damped Newton dynamics until forces became less
than $0.01$ eV/\AA. All the calculations were performed using the supercell
approach. Details of the cell shape and size, as well as {\bf k}-point
sampling, for the different geometries considered, are given along with the
results in the following sections.

\section{Results}\label{res}

\subsection{Bulk and Surface Properties}\label{res_clean}

In order to assess the validity of our approach, and for completeness, we
first determined the lattice constant of the bulk material as well as the
properties of the clean Pt(111) surface --- surface energy, relaxation,
excess surface stress, and work function. As a reminder, the surface stress
tensor, $g_{\alpha\beta}$, is given by
   \begin{equation}
   g_{\alpha \beta} = \gamma \delta_{\alpha \beta} +
                          d \gamma / d \varepsilon_{\alpha \beta}
   \end{equation}
where $\gamma$ is the surface energy per unit area,
$\varepsilon_{\alpha\beta}$ is the strain tensor and $\delta_{\alpha\beta}$
is the Kronecker delta function. The second term in this equation represents
the excess surface stress. For the clean surface, we used a ($1\times1$)
supercell consisting of 5 or 7 (111) atomic planes plus $\sim$10 \AA\ of
vacuum. Integration of the first Brillouin zone was done over a uniform grid
of 100 {\bf k} points in the $x-y$ plane, which was found to yield
well-converged results, e.g., within 2 meV for the surface energy. The latter
was calculated by comparing to a bulk-like ($1\times1$) supercell containing
3 layers (and of course no vacuum). The same {\bf k}-point density was used
for the $x-y$ plane; to compensate for the smaller size of the cell in the
$z$ direction, the two-dimensional grid was replicated 4 times along $z$ so
as to get a density of points similar to that in the $x-y$ plane. For the
excess surface stress, we varied the in-plane lattice constant for both the
clean surface and the bulk, while keeping the atoms' $z$ coordinates fixed to
their bulk-like values, and examined the concomitant variations in the total
energy.

The results are listed in Table \ref{tests}, along with those from other {\em
ab-initio} calculations and experimental values when available. Evidently,
our LDA lattice constant is consistent with previous calculations and with
experiment. The GGA value, in contrast, while in agreement with other
calculations, overestimates somewhat the lattice constant, by more than
$2\%$; it is a well-known fact that the GGA yields larger lattice constant
than the LDA\cite{khein} --- actually overcompensates in the present case.
For the surface energy and excess surface stress, our LDA results agree well
with previous calculations. We note that the excess surface stress is large
and positive, meaning that the Pt(111) surface is under significant tensile
stress, i.e., would prefer a smaller lattice constant. Unfortunately, to our
knowledge, there exists no experimental determination of these quantities. In
the GGA, both the surface energy and the excess surface stress decrease. This
effect of the GGA on the surface energy was actually predicted from jellium
calculations.\cite{perdew}

We have also calculated the top and second layer relaxation, $\Delta d_{12}$
and $\Delta d_{23}$, i.e., the change in interlayer spacing relative to the
bulk value. We find a small (0.4$\%$) outward relaxation for the top layer.
This is quite a bit smaller than the value reported in Ref.\
\onlinecite{feibelman1}, 1.25$\%$. The origin of the discrepancy between the
two LDA calculations is not clear; it may be due to different choices of
basis sets (LCAO in Ref.\ \onlinecite{feibelman1} versus plane waves here).
The GGA, interestingly, leads to a small {\em inward} relaxation. Currently
available experimental data vary widely, in the range 0--2.5$\%$, and are
therefore not of much help in resolving the issue. For the second layer,
theory and experiment agree that it should be insignificant, i.e., $\Delta
d_{23}$ is a small fraction of a percent.

For the work function, finally, the experimental values also vary quite a
bit, in the range 5.77--6.10 eV. According to Kaack and Fick,\cite{kaack}
however, the work function has a small temperature dependence, decreasing
slightly with temperature. Since our calculations are performed at 0~K, we
expect that they should compare well with the largest experimental values. We
find, indeed, that the LDA result, 6.07 eV, is in excellent agreement with
the largest experimental number, 6.10 eV. The GGA value, in contrast, is
significantly smaller --- 5.70 eV --- indicating, once more, that the LDA
provides a better description of Pt than the LDA. In spite of this, the two
approximations will carefully be examined in the context of SFE and diffusion
barriers.

\subsection{Step Formation Energy}\label{res_vic}

We come now to the heart of the matter, namely the energetics of step
formation. In order to determine the SFE, we constructed vicinal surfaces
(using rectangular surface cells) appropriate to each type of steps. For the
(100)-faceted step, we examined both a (211) and a (332) surface; the former
has 3 atoms per terrace while the latter has 5. For the (111)-faceted step,
only the (221) surface, which contains 4 atoms per terrace, is considered.
Again, here, a vacuum region of approximately 10 \AA\ was included in all
cases.

The energies of the vicinal surfaces were determined using the same procedure
as in the case of the clean surface, i.e., by comparing to an appropriate
bulk model. In order to minimize the error arising from the use of different
geometries, the bulk reference system for a given vicinal surface was always
taken to have the same in-plane geometry as the surface. Thus, the same {\bf
k} points were used in the $x-y$ plane, while for the $z$ coordinate, the
grid was adjusted to yield a comparable density. In total, the bulk
supercells corresponding to the (211), (322), and (221) surfaces contained 6,
34, and 18 atoms, respectively.

The SFE is given, simply, by the difference in energy between a surface with
a step and one without. Since we are dealing with vicinal surfaces here, this
is equivalent to subtracting from the vicinal-surface energy (per terrace),
$\sigma_{\rm vic}$, that portion of the clean-surface energy corresponding to
the exposed (111) area.\cite{feibelman0} If we call $\sigma_{(111)}$ the
clean-(111)-surface energy per atom and neglect step-step interactions we
find, for the (100)-faceted step:
   \begin{equation}
   E^{\rm SF}_{(100)} = \sigma_{\rm vic} - ( N - \frac{1}{3} ) \sigma_{(111)},
   \label{step-100}
   \end{equation}
and for the (111)-faceted step:
   \begin{equation}
   E^{\rm SF}_{(111)} = \sigma_{\rm vic} - ( N - \frac{2}{3} ) \sigma_{(111)},
   \label{step-111}
   \end{equation}
where $N$ is the number of atoms per terrace, as defined earlier.

It is evident from Eqs.\ \ref{step-100} and \ref{step-111} that, in order to
determine reliably the SFE ratio, {\em very} accurate surface energies are
required for both vicinal and clean surfaces. It is our purpose here to
assess carefully the accuracy of our calculations through a detailed
convergence study. As explained earlier, the error arising from the supercell
geometry is minimized by always comparing surface and bulk energies obtained
using the same in-plane periodicity and {\bf k}-point density. Of course, it
is essential that the energies be converged with respect to Brillouin-zone
integration; for the clean (111) surface, the {\bf k}-point sampling scheme
used here leads to values converged within 2 meV, as discussed in Sec.\
\ref{res_clean}. For the vicinal surfaces, we used a similar sampling scheme
and, as we will see below, the error is of the order of a few meV, so that
differences in energy (e.g., between steps) of a few hundredths of an eV {\em
are} significant.

The results for the (100)-faceted step under a variety of theoretical
conditions are listed in Table \ref{step100}. First, we examine the effect of
relaxation and size within the LDA. With all atoms in the bulk-like
configuration (referred to as ``rigid'' in Table \ref{step100}), we find the
SFE to be rather insensitive to the size of the supercell: adding two layers
to the (211) slab increases the SFE by a mere 0.02 eV/(step atom), while
using a (322) surface instead of a (211) leads to a small decrease of 0.01
eV/(step atom). (The number of layers refers to the number of (111)-like
layers in the slab before a rotation is applied to make the surface vicinal.)
However, upon relaxing {\em all} atoms, except those in the central (111)
layer (configurations referred to as ``relaxed'' in Table \ref{step100}), the
SFE is found to decrease strongly, from 0.62 to 0.43 eV/(step atom) for the
(322) surface. For the relaxed configurations, we have also examined the
convergence with respect to the {\bf k}-point sampling. In all cases, the SFE
changes by at most 0.01 eV/(step atom) upon increasing the number of {\bf k}
points. Thus, we estimate the SFE for the (100)-faceted step to be
0.43$\pm$0.02 eV/(step atom) within the LDA.

For the (111)-faceted step, in view of the above results, we have studied a
single vicinal surface, namely the (221), at fixed and converged {\bf
k}-point density, as indicated in Table \ref{step111}. Again, here,
relaxation affects strongly the SFE, which decreases very markedly --- from
0.64 to 0.38 eV for a 5-layer slab in the LDA. However, increasing the
thickness from 5 to 7 layers brings about no significant changes in the SFE.
Thus, our best LDA-SFE value for the (111)-faceted step is 0.38$\pm$0.02
eV/(step atom).

Our calculations indicate, therefore, that the (111)-faceted step has a lower
formation energy than the (100)-faceted step --- 0.38 versus 0.43 eV (in the
LDA), leading to a ratio $E^{\rm SF}_{(111)} / E^{\rm SF}_{(100)}$ of
0.88$\pm$0.07. This is in excellent agreement with the experimental value of
0.87$\pm$0.02,\cite{michely} but at variance with a previous LDA calculation
by Feibelman,\cite{feibelman0} who found that the two steps are nearly
equivalent, that is, 0.46 eV for the (111)-faceted step versus 0.47 eV for
the (100), i.e., a ratio of 0.98. The SFE values differ from Feibelman's not
only in a relative sense, but also in an absolute sense: the values we find
are significantly smaller, by 0.08 eV/(step atom) for the (111)-faceted step
and 0.04 eV/(step atom) for the (100)-faceted step. Though the reasons for
these differences are not clear, they may originate in the choice of basis
sets: while we use plane waves, Feibelman employs Gaussian orbitals which are
more sensitive to the details of the atomic configuration, as discussed in
Ref.\ \onlinecite{feibelman0}. In view of this, it might perhaps be the case
that a Gaussian basis set lacks the accuracy needed to resolve such small
energy differences as those involved here.

A value of 0.37 eV/(step atom) for the (111)-faceted SFE has also been
estimated from the experimental surface free-energy anisotropy between the
(110) and (111) surfaces.\cite{bonzel} While this corresponds quite closely
to our value of 0.38 eV/(step atom), the agreement is fortuitous since the
above result was obtained assuming a value of the surface energy of 0.097
eV/\AA $^2$, much lower than that calculated here, 0.124 eV/\AA $^2$. [Using
the latter value for the surface energy would lead to a (111)-faceted SFE
0.45 eV/(step atom) in the approach of Ref.\ \onlinecite{bonzel}.]

Recently, some concerns have been expressed regarding the procedure used here
to determine the surface energy, which should diverge as the thickness of the
slab increases.\cite{boettger,fiorentini} We reinterpreted our results using
the approach suggested in Ref.\ \onlinecite{fiorentini} and found only small
changes in the (111) surface energy, now 0.121 eV/\AA$^2$ rather than 0.124
eV/\AA$^2$; for the SFE, we obtain now 0.41 and 0.46 eV/(step atom) for the
(111)- and (100)-faceted step, respectively, compared to 0.38 and 0.43
eV/(step atom) using the usual approach. The SFE ratio remains approximately
unchanged, 0.89 versus 0.88. We are thus led to conclude that, while the
uncertainty on the SFE might be of the order a 0.03 eV, the value we find for
the ratio is accurate to a few percent, and is not affected by the numerical
procedure used.

As mentioned in the previous section, the LDA seems to provide a better
description of bulk Pt, as well as of the (111) surface, than the GGA. In
view of this, it is expected that the vicinal surfaces are also better
represented within the LDA. The question remains open, however, because there
exists no firm experimental data to compare our results to, and it is
therefore of interest to calculate the SFE also within the GGA. The results
are given in Tables \ref{step100} and \ref{step111}. We observe the GGA-SFE
to be systematically lower than the corresponding LDA values, as is also true
of the (111) surface energy. Further, convergence with respect to both size
and {\bf k}-point density is similar in the two approximations. We therefore
conclude to GGA values of 0.29$\pm$0.02 and 0.25$\pm$0.02 eV/(step atom) for
the (100)- and (111)-faceted step, respectively. The resulting SFE ratio is
0.86$\pm$0.10, essentially unchanged from the LDA value, namely
0.88$\pm$0.07.

\begin{figure}
%\vspace*{-0.25in}
\epsfxsize=3.in \epsfbox{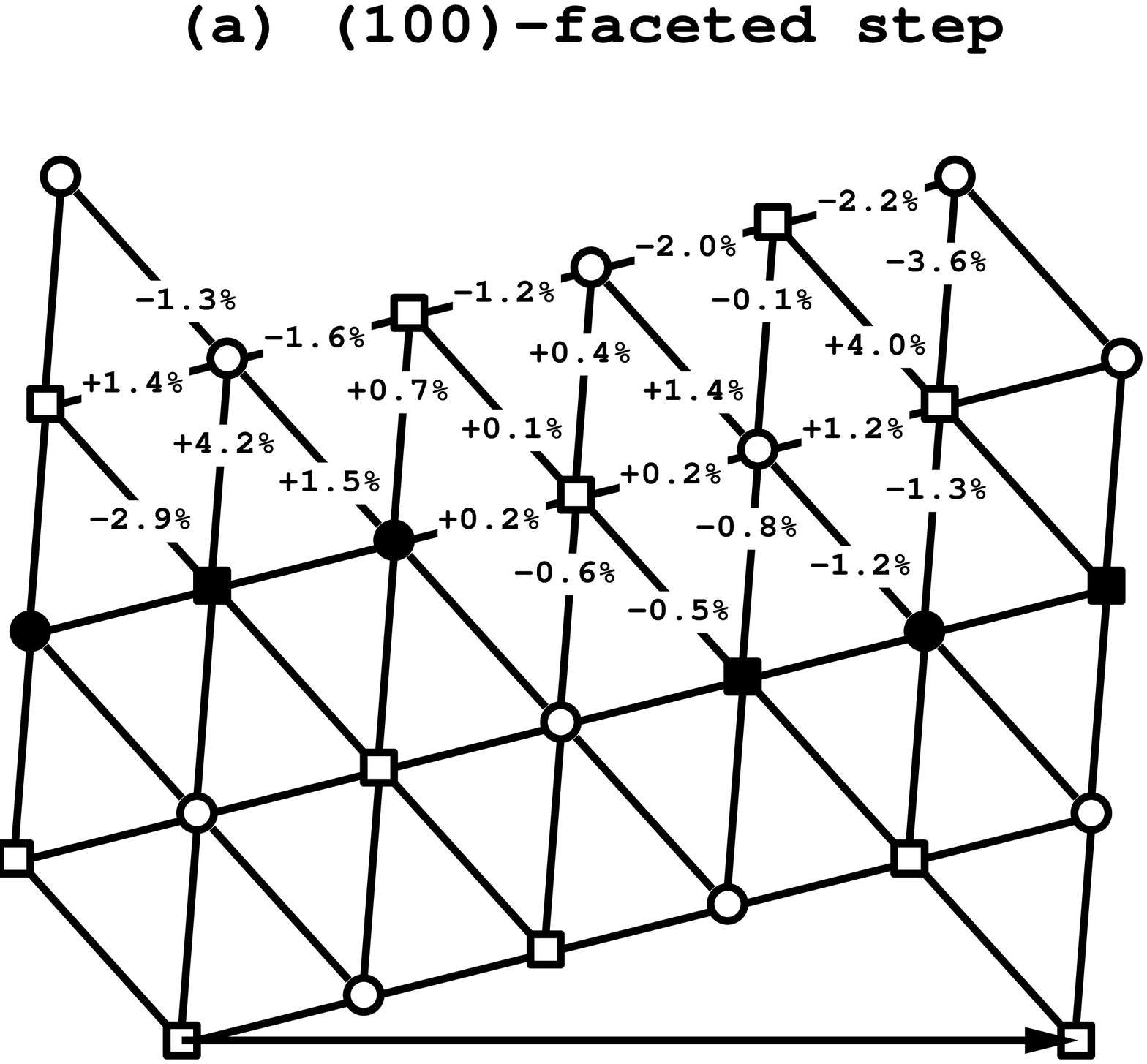}
%\vspace{-0.75in}
\vspace*{0.1in}
\epsfxsize=3.in \epsfbox{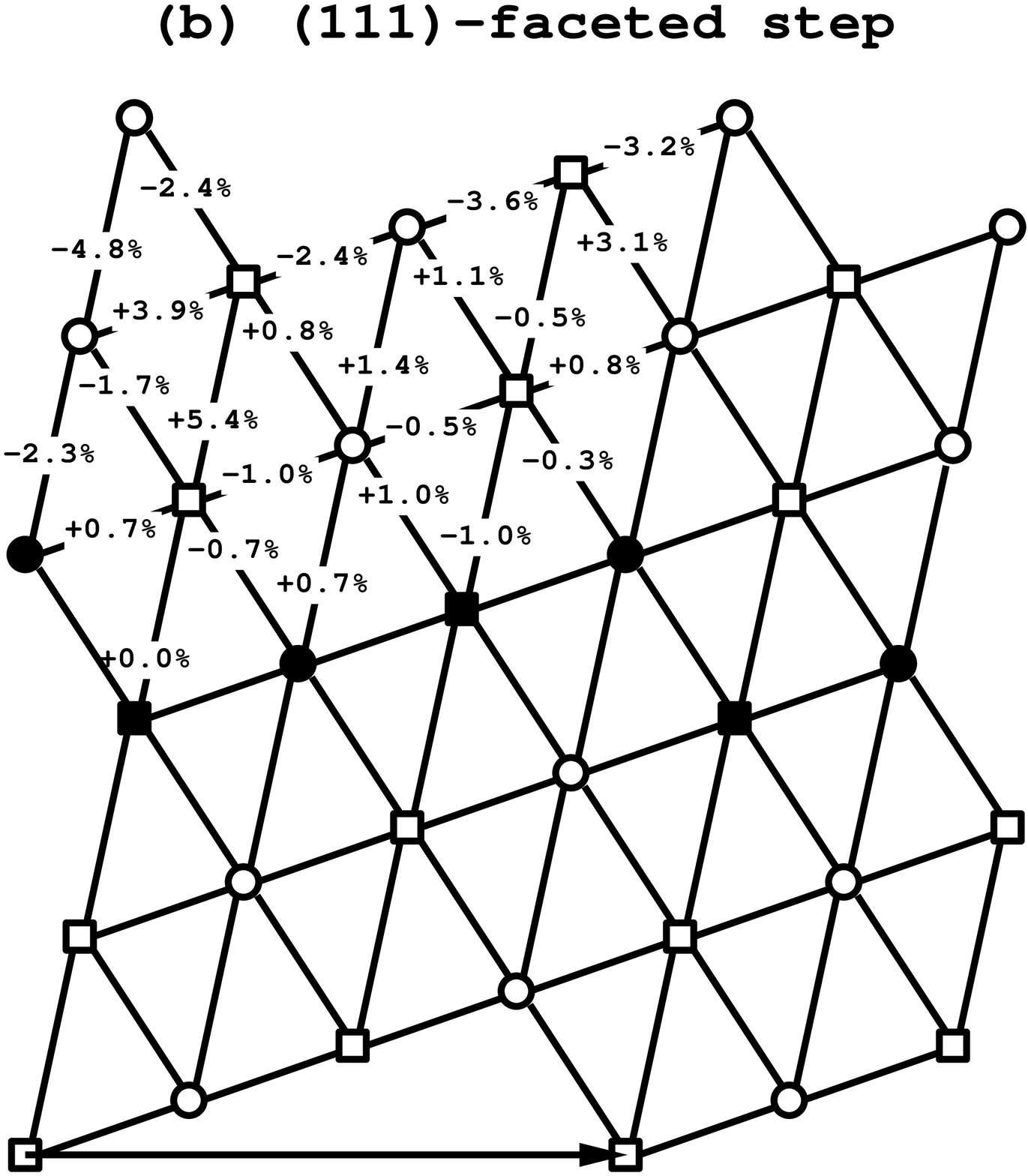}
\vspace{0.25in}
\caption{
Side view of the supercell showing the relaxed geometry for (a) the
(100)-faceted step [as obtained from the (322) surface] and (b) the
(111)-faceted step [from the (221) surface]. The filled symbols represent the
middle (111) layer, which is fixed in its bulk-like position. The circles are
in the plane of the sheet while the squares are in a different plane, distant
by $d/2$, where $d$ is the nearest-neighbor distance. The arrows indicate the
periodicity of the supercell.
\label{rel_geo}
}
\end{figure}

In order to understand why the two steps have different formation energies,
it is of interest to consider, first, the (111)- to (100)-faceted SFE ratio
in the {\em unrelaxed} (bulk-like) configuration. We find, from Tables
\ref{step100} and \ref{step111}, this ratio to be equal to 1.03$\pm$0.07
(using the LDA), compared to about 0.88 for the relaxed configurations, as we
have seen above. Thus, before the atoms relax, the two steps are nearly
equivalent [with perhaps a slight preference for the (100)-faceted step], as
could be expected from a simple nearest-neighbor model as explained in the
Introduction. Evidently, therefore, the observed step anisotropy is closely
related to relaxation, and this can be understood in the following way: As we
have seen in Sec.\ \ref{res_clean}, the Pt(111) surface is under large
tensile stress, which can be locally relieved at steps. However, because the
atomic configurations are different (albeit slightly) for the two kinds of
steps, the relaxation patterns also differ, and lead to different energetics.
This can in fact be seen very clearly in Fig.\ \ref{rel_geo}, where we plot
the displacement patterns for the two types of steps: Some atoms suffer very
large displacements --- by as much as a few percent (relative to the bulk
nearest-neighbor distance) for those that sit closest to the steps. More
important, it is also clear from this figure that the displacements
associated to the (111)-faceted step are larger than for the (100)-faceted
step, i.e., the former can relieve stress more efficiently than the latter,
and is thus energetically more favorable. It should be mentioned that the
displacements we find here differ from Feibelman's\cite{feibelman0} by as
much as 1$\%$ in some cases, and might possibly explained the discrepancy
between the two sets of results; this is likely related, again, to different
choices of basis functions.

The relation between relaxation and stress can be understood in a more
quantitative manner by considering the change in energy resulting from the
displacement inwards of an edge atom, i.e., in the direction normal to the
step and parallel to the terrace. Starting with both step models in their
bulk-like geometry and moving an edge atom by the same amount for the two
steps, we find that the (111)-faceted step, because of its triangular
geometry, releases more energy than the (100)-faceted step: for a
displacement of 0.14 \AA, corresponding approximately to the observed
relaxation, we find the (111)-faceted step to be already 22 meV lower in
energy than the (100)-faceted step.

Clearly, therefore, the difference in SFE arises from the large excess
surface stress of the Pt(111) surface --- 0.25 eV/\AA$^2$ (cf.\ Table
\ref{tests}), about twice as large as the surface energy, 0.124 eV/\AA$^2$.
We may compare this with the corresponding situation for Ir/Ir(111), where
the equilibrium island shape is nearly hexagonal.\cite{fu} In this case, the
excess surface stress is 0.128 eV/\AA$^2$,\cite{needs} significantly smaller
than the surface energy, 0.204 eV/\AA$^2$. Evidently, large surface energies
lead to large SFE, and large excess surface stresses to large differences
between the two steps. These observations therefore suggest that, as a ``rule
of thumb'', the SFE ratio should differ from one (i.e., non-hexagonal
equilibrium island shape) when the excess surface stress is sizeably larger
than the surface energy. This is in fact the case of Al(111)\cite{needs2} and
Au(111),\cite{needs} while the opposite is true of Rh\cite{filippetti} and
Cu.\cite{boisvert2} To our knowledge, no information on the equilibrium
island shape is available for Rh and Cu, but we would predict it to be
hexagonal in both cases. For Au, reconstruction\cite{barth} is likely to be
important in determining the island shape. For Al, finally, {\em ab initio}
calculations of the kind presented here have been performed by Stumpf and
Scheffler,\cite{stumpf1} and a SFE ratio of 0.93 is indeed found. In this
case, the ratio does not seem to be affected by relaxation, contrary to our
results for Pt, but different electronic orbitals are involved --- $sp$ for
Al and $d$ for Pt.

There have been other calculations of the equilibrium island shape on
Pt(111). Using a tight-binding model, Papadia {\em et al.\ } found
essentially no difference between the two steps in their bulk-like
configuration, as is the case here, but did not consider
relaxation.\cite{papadia} Within equivalent-crystal theory, Khare and
Einstein found a ratio close to unity (0.968) for the SFE, without allowing
in-plane relaxation;\cite{khare} also, in this approach, the surface energy
is predicted to be 0.076 eV/\AA $^2$, quite a bit smaller than our 0.124
eV/\AA $^2$. Fully-relaxed calculations using the semi-empirical
embedded-atom method (EAM) have also been performed and lead to a SFE ratio
very close to one;\cite{nelson} it is however doubtful that the EAM potential
is robust enough to account for the small energy differences involved here.
Other approaches have been proposed, based on coordination- or
orientation-dependent bonds,\cite{fallis,barkema} which do not take
relaxation effects into account.

We also list, in Tables \ref{step100} and \ref{step111}, the work function
for the different surfaces examined. As expected (see for instance Ref.\
\onlinecite{meth92}), and already observed by Feibelman,\cite{feibelman0} the
work function is smaller for the vicinal surfaces than for the (111) surface.
We also observe, in the case of the (100)-faceted step, a small dependence on
the the terrace length. This agrees with Feibelman's calculations, while a
stronger dependence is reported from experiment.\cite{besocke} This might be
due to the fact that terraces studied in experiment are much wider than ours
and/or surface contamination.\cite{feibelman0}

\subsection{Atom Adsorption and Diffusion}\label{res_ad}

We now discuss adsorption and diffusion of a Pt atom on the Pt(111) surface.
For these calculations, an adatom is added on one surface of the Pt slab
while the other surface is constrained to its bulk-like configuration. In
order to determine the energies at the two adsorption sites as well as the
barrier for diffusion, we considered both a ($2\times2$) and a ($3\times3$)
cell with, again, approximately 10 \AA\ of vacuum. Unless otherwise noted,
the integration over reciprocal space was performed using a mesh of 16
equidistant {\bf k} points for the ($2\times2$) cell and 9 {\bf k} points for
the ($3\times3$) cell.

First, starting with the ($2\times2$) system, we examined convergence with
respect to the number of layers, which we varied from 3 to 6. The results are
given in Table \ref{barrier}. We observe significant changes upon going from
3 to 4 (for a given state of relaxation), while increasing this number
further brings about changes of at most 0.01 eV; thus, 4 layers seem to be
sufficient for reliable estimates of both the diffusion barrier, $E_{\rm d}$,
and the difference in adsorption energies, $\Delta E_{ads}$.

The barrier we obtain for the {\em unrelaxed} substrate, 0.41 eV, and the
difference in adsorption energies, 0.10 eV, agree well with previous LDA
calculations.\cite{mortensen,feibelman1} It is however clear from Table
\ref{barrier} that relaxation effects are again here important: If we allow
the topmost layer to relax, $E_{\rm d}$ decreases to 0.34 eV and $\Delta
E_{\rm ads}$ increases to 0.17 eV; including second-layer relaxation as well
results in relatively minor changes to the energies (less than 0.02 eV).

It should be noted that, because of the asymmetry between the two adsorption
sites, there are in fact two barriers for diffusion. However, the difference
between the two barriers is such that, for temperatures of interest, it is
the highest-energy barrier that limits diffusion, i.e., the adatom will
escape rapidly from the low-energy adsorption state but get trapped in the
high-energy site.

We have also examined convergence with respect to {\bf k}-point sampling,
energy cutoff, and lateral size. As indicated in Table \ref{barrier}, we find
in all cases very modest changes of at most 0.01 eV. Likewise, using the GGA
does not lead to appreciable changes to the energy barrier, while $\Delta
E_{\rm ads}$ decreases by about 0.03 eV. For the reasons discussed in Sec.\
\ref{res_clean}, we suspect that the LDA values are more accurate; our best,
highly-converged estimates of $E_{\rm d}$ and $\Delta E_{\rm ads}$ are thus
0.33$\pm$0.03 and 0.17$\pm$0.03 eV, respectively. As discussed in Ref.\
\onlinecite{mortensen}, the large difference between the fcc and the hcp site
is related to the angular character of the $d$ orbitals, whose bonding
strength depends on the filling and radial quantum number of the bands. Due
to this difference, the transition site for jump diffusion does not lie {\em
exactly} midway between the two equilibrium sites but, rather, about 0.07
\AA\ towards the hcp site. The potential energy surface is however very flat
in this region, changing by no more than 0.01 eV upon going from the
transition to the midpoint site.

There is very little experimental information available for $\Delta E_{\rm
ads}$; as mentioned in the Introduction, only a lower limit of 0.06 eV has
been determined,\cite{golzhauser} and this is consistent with our results.
For the diffusion barrier, a value of 0.25$\pm$0.02 eV has been inferred from
field-ion microscopy (FIM) measurements in the temperature range 92--100
K.\cite{feibelman1} Also, based on a comparison between STM measurements of
the island density and kinetic Monte Carlo simulations between 110 K and 160
K, an estimate of 0.26$\pm$0.01 eV is obtained.\cite{bott} Both values agree,
but disagree with our result of 0.33$\pm$0.03 eV, which is surprising in view
of the high-level of accuracy and convergence of our calculations.

One possible explanation for this disagreement would be the neglect, in our
calculations, of dynamical effects. These cannot be assessed directly from
first-principles but have been shown to be insignificant at low temperatures,
either from the calculation of dynamical corrections to the transition state
theory\cite{cohen} or from empirical molecular-dynamics
simulations.\cite{boisvert3,ferrando,kallinteris} Experimentally, it is clear
that the FIM value\cite{feibelman1} suffers from poor statistics --- only two
points are used to determine the Arrhenius parameters --- and therefore the
error bar is large. Concerning the STM experiment, it has been pointed out by
the authors that small-cluster mobility could affect the Monte-Carlo estimate
of the barrier if its energy is close to that for adatom diffusion, in which
case the quoted value would be a lower bound to the actual barrier. As a
final point, it should be mentioned that many different empirical potentials
have been used to determine the diffusion barrier of Pt adatoms on
Pt(111),\cite{feibelman1,majerus,li,liu,stoltze,basset} yielding to values in
the range 0.01--0.18 eV, i.e., much lower than the experimental value, which
we argue, is a lower bound to the actual barrier. Thus, such models are
clearly too crude to provide a proper description of the energetics of
diffusion for the present system.

\section{Summary}\label{conc}

We have used highly-accurate {\em ab-initio} methods to calculate the ratio
of (111)- to (100)-faceted step formation energies on the (111) surface of
Pt. We find, in excellent agreement with experiment, (111)-faceted steps to
be favoured over (100) in a ratio of about 0.88; as a consequence, the island
on this surface should be clearly non-hexagonal. The difference between the
two steps is related to the large tensile stress of the Pt(111) surface,
which is released in a different manner because of differences in the local
topology. Our calculations underline the importance of relaxation in such
cases: while the two steps are about equivalent for the unrelaxed substrate,
relaxation does bring about large changes in the formation energies.
Likewise, relaxation is important to a proper determination of
equilibrium-site energies on the flat (111) surface. We find the fcc site to
be preferred over the hcp by a sizeable 0.17 eV (after full relaxation),
consistent with experiment which provides a lower bound of 0.06 eV for the
difference between the two sites.

We have also calculated the energy barrier for adatom diffusion and found a
fully-relaxed value of 0.33 eV. While constituting an improvement over
previous calculations, this value remains larger than experiment, by about
0.07 eV. The discrepancy might be due to limitations of our theoretical
approach (e.g., finite size), but it might also due to errors in the
interpretation of the experimental data --- poor statistics, neglect of
small-cluster contributions (e.g., dimers) to mass transport, i.e., incorrect
assumption regarding the critical nucleus size. More experiments are needed
to clarify this point. Likewise, it would be of interest that measurements of
the SFE and difference in energy between fcc and hcp sites be carried out so
as to assess the validity of LDA (versus GGA) in the present context.

\acknowledgements

G.B.\ acknowledges the warm hospitality of the Theory Department of the
Fritz-Haber-Institut where a large part of this work was performed. We are
grateful to Martin Fuchs for help with generating the pseudopotentials and to
Alexander Kley, Christian Ratsch, Paolo Ruggerone, Ari P Seitsonen, and Byung
Deok Yu for stimulating discussions. This work was supported by grants from
the Natural Sciences and Engineering Research Council (NSERC) of Canada and
the ``Fonds pour la formation de chercheurs et l'aide {\`a} la recherche''
(FCAR) of the Province of Qu{\'e}bec. One of us (G.B.) is thankful to NSERC and
FCAR for financial support. We are grateful to the ``Services informatiques
de l'Universit{\'e} de Montr{\'e}al'' for generous allocations of computer
resources. Part of the calculations reported here were carried out on the
IBM/SP-2 at the CACPUS (``Centre d'applications du calcul parall{\`e}le de
l'Universit{\'e} de Sherbrooke'').

\newpage
\onecolumn
\widetext

\begin{center}
\begin{table}
\caption{
Bulk and clean (111) surface properties of platinum: lattice constant $a$,
surface energy $\sigma$, excess surface stress $\tau$, top and second layer
relaxation, $\Delta d_{12}$ and $\Delta d_{23}$, and work function $W$.
}
\label{tests}
\begin{tabular}{lcccccc}
& $a$ & $\sigma$ & $\tau$ & $\Delta d_{12}$ & $\Delta d_{23}$ & $W$  \\
&(\AA)&(eV/\AA$^2$)&(eV/\AA$^2$)&($\% d_{\rm bulk}$)&($\% d_{\rm bulk}$)&(eV)\\
\tableline
Present LDA & 3.92 & 0.124 & 0.25 & 0.4 & $-$0.2 & 6.07 \\
Present GGA & 4.00 & 0.097 & 0.22 & $-$0.4 & 0.0 & 5.70 \\
Other LDA   & 3.87,\cite{needs} 3.89,\cite{feibelman1,ozolins,feibelman2}
              3.90\cite{khein}&0.137\cite{needs,feibelman2}&
              0.213,\cite{needs} 0.289\cite{feibelman2}&1.25\cite{feibelman1} &
              $-$0.05\cite{feibelman1} & 6.10\cite{feibelman1} \\
Other GGA   & 3.97\cite{ozolins,khein}& & & & & \\
expt.       & 3.91\cite{touloukian}& & &
              $< \mid 2.5 \mid$,\cite{kesmodel,hayek}
              $< \mid 2 \mid$,\cite{bogh} 1.5$\pm$1.0,\cite{vanderveen}
              & 0.0\cite{barbieri} &
              5.77--6.10\cite{kiskinova,salmeron,derry,alrot,cassuto,kaack} \\
              & & & & 1.1$\pm$0.5,\cite{barbieri}
               1.0,\cite{adams}  $< \mid 0.4 \mid$,\cite{davies}  & & \\
\end{tabular}
\end{table}
\end{center}

\begin{center}
\begin{table}
\caption{
Step formation energy (SFE) for the (100)-faceted step on Pt(111) (in eV per
step atom) and work function $W$ for the corresponding vicinal surfaces (see
text), as a function of the number of (111) layers, $N_L$, number of {\bf k}
points, $N_{k}$, and approximation scheme for the exchange-and-correlation
energy (XC).
}
\label{step100}
\begin{tabular}{lccccc}
surface       & $N_L$ & $N_{k}$ &   XC    & SFE    &  $W$  \\
              &       &         &         & (eV/at)&  (eV) \\ \tableline
(211)-rigid   &    5  &     8   &   LDA   & 0.63   &  5.88 \\
(211)-relaxed &    5  &     8   &   LDA   & 0.45   &  5.92 \\
(211)-relaxed &    5  &    12   &   LDA   & 0.46   &  5.93 \\
(211)-rigid   &    7  &     8   &   LDA   & 0.65   &  5.89 \\
(211)-relaxed &    7  &     8   &   LDA   & 0.47   &  5.94 \\
              &       &         &         &        &       \\
(211)-rigid   &    5  &     8   &   GGA   & 0.50   &  5.58 \\
(211)-relaxed &    5  &     8   &   GGA   & 0.34   &  5.63 \\
(211)-relaxed &    5  &    12   &   GGA   & 0.34   &  5.64 \\
(211)-rigid   &    7  &     8   &   GGA   & 0.52   &  5.59 \\
(211)-relaxed &    7  &     8   &   GGA   & 0.35   &  5.63 \\
              &       &         &         &        &       \\
(322)-relaxed &    5  &     4   &   LDA   & 0.44   &  5.96 \\
(322)-rigid   &    5  &     8   &   LDA   & 0.62   &  5.94 \\
(322)-relaxed &    5  &     8   &   LDA   & 0.43   &  5.95 \\
              &       &         &         &        &       \\
(322)-rigid   &    5  &     4   &   GGA   & 0.46   &  5.64 \\
(322)-relaxed &    5  &     4   &   GGA   & 0.29   &  5.66 \\
\end{tabular}
\end{table}
\end{center}

\begin{center}
\begin{table}
\caption{
Same as Table II but for the (111)-faceted step.
}
\label{step111}
\begin{tabular}{lccccc}
surface       & $N_L$ & $N_{k}$ &   XC    & SFE    &  $W$  \\
              &       &         &         & (eV/at)&  (eV) \\ \tableline
(221)-rigid   &    5  &     8   &   LDA   & 0.64   &  5.84 \\
(221)-relaxed &    5  &     8   &   LDA   & 0.38   &  5.80 \\
(221)-relaxed &    7  &     8   &   LDA   & 0.38   &  5.82 \\
              &       &         &         &        &       \\
(221)-rigid   &    5  &     8   &   GGA   & 0.50   &  5.54 \\
(221)-relaxed &    5  &     8   &   GGA   & 0.25   &  5.50 \\
\end{tabular}
\end{table}
\end{center}

\begin{center}
\begin{table}
\caption{
Adatom diffusion barrier $E_{\rm d}$ and difference in adsorption energies
between fcc and hcp sites, $\Delta E_{\rm ads} = E_{\rm ads}^{\rm fcc} -
E_{\rm ads}^{\rm hcp}$, for Pt on Pt(111) under various calculational
conditions, as discussed in the text.
}
\label{barrier}
\begin{tabular}{lccccc}
XC&supercell&allowed to relax&{\bf k} point grid & $E_{\rm d}$ & $\Delta E_{\rm ads}$ \\
   &         &       &    & (eV)      &   (eV)       \\ \tableline
LDA&(2$\times$2), 3-layer&adatom only       & 4$\times$4   & 0.47 &  0.22 \\
LDA&(2$\times$2), 3-layer&adatom + top layer& 4$\times$4   & 0.41 &  0.26 \\
   &                     &                  &       &      &       \\
LDA&(2$\times$2), 4-layer&adatom only       & 4$\times$4   & 0.41 &  0.10 \\
LDA&(2$\times$2), 4-layer&adatom + top layer& 4$\times$4   & 0.34 &  0.17 \\
GGA&(2$\times$2), 4-layer&adatom + top layer& 4$\times$4   & 0.33 &  0.14 \\
LDA&(2$\times$2), 4-layer&adatom + 2 top layers & 4$\times$4 & 0.36 &  0.17 \\
LDA&(2$\times$2), 4-layer&adatom + top layer    & 5$\times$5   & 0.35 &  0.17 \\
LDA(50 Ry)&(2$\times$2), 4-layer&adatom + top layer&4$\times$4&0.35&      \\
   &                     &                  &       &      &       \\
LDA&(2$\times$2), 5-layer&adatom + top layer& 4$\times$4   & 0.33 &       \\
   &                     &                  &       &      &       \\
LDA&(2$\times$2), 6-layer&adatom + top layer& 4$\times$4   & 0.35 &       \\
   &                     &                  &       &      &       \\
LDA&(3$\times$3), 4-layer&adatom + top layer& 3$\times$3   & 0.33 &       \\
   &                     &                  &       &      &       \\
\end{tabular}
\end{table}
\end{center}

\newpage
\twocolumn
\narrowtext

\input{epsf.tex}

%\vfill
\vspace{1.0cm}

\vfill

\end{document}